\documentclass[floatfix,aps,prl,twocolumn,superscriptaddress,noeprint]{revtex4-1}

\usepackage{graphicx}
\usepackage{amsmath}
\usepackage{amssymb}
\usepackage{booktabs}
\usepackage{xcolor}

\usepackage[utf8]{inputenc}
\usepackage{natbib}

\usepackage{bm}
\usepackage{tabularx}   
\usepackage{MnSymbol,wasysym}

\begin{document}

\definecolor{blueNico}{rgb}{0.36, 0.54, 0.66}
\newcommand{\Nico}[1]{{\textcolor{blueNico}{#1}}}
\newcommand{\Jeff}[1]{{\textcolor{red}{#1}}}
\newcommand{\Derek}[1]{{\textcolor{orange}{#1}}}

\title{Disorder suppresses chaos in viscoelastic flows}

\author{Derek~M. Walkama}
\affiliation{Department of Mechanical Engineering, Tufts University, 200 College Avenue, Medford, Massachusetts 02155, USA}
\affiliation{Department of Physics and Astronomy, Tufts University, 574 Boston Avenue, Medford, Massachusetts 02155, USA}
\author{Nicolas Waisbord}
\affiliation{Department of Mechanical Engineering, Tufts University, 200 College Avenue, Medford, Massachusetts 02155, USA}
\author{Jeffrey~S. Guasto}
\affiliation{Department of Mechanical Engineering, Tufts University, 200 College Avenue, Medford, Massachusetts 02155, USA}


\begin{abstract}
Viscoelastic flows transition from steady to time-dependent, chaotic dynamics under critical flow conditions, but the implications of geometric disorder for flow stability in these systems are unknown. 
Utilizing microfluidics, we flow a viscoelastic fluid through arrays of cylindrical pillars, which are perturbed from a hexagonal lattice with various degrees of geometric disorder. 
Small disorder, corresponding to $\sim10\%$ of the lattice constant, delays the transition to higher flow speeds, while larger disorders exhibit near-complete suppression of chaotic velocity fluctuations.
We show that the mechanism facilitating flow stability at high disorder is rooted in a shift from extension-dominated to shear-dominated flow type with increasing disorder.
\end{abstract}


\maketitle

Viscoelastic fluids encompass a wide range of complex materials having a mechanical response to strain that lies between elastic solids and viscous fluids~\cite{Chen2010}.
Even in the absence of inertia, viscoelastic flows spontaneously become time dependent when elastic stresses overcome viscous stresses.
This condition is often characterized by a Weissenberg number, $\mathrm{Wi}=\tau \dot{\gamma}>\mathrm{Wi}_\mathrm{cr}\sim 1$, where $\dot{\gamma}$ is the typical shear rate and $\tau$ is the fluid relaxation time~\cite{Larson1992}.
Purely elastic instabilities manifest as spatio-temporal chaotic flow and elastic turbulence~\cite{Groisman2000, Groisman2004} in a wide range of natural and industrial applications: Elasticity generates secondary flows of DNA and blood suspensions in biological systems~\cite{Gulati2008, Thiebaud2014}, hydrodynamic resistance increases~\cite{Denn2004} along with power consumption and cost in polymer processing, and elastic instabilities enhance mixing and dispersion in microfluidic and porous media flows~\cite{Groisman2001, Clarke2015,Scholz2014}. 
Experimental \cite{Muller1989, McKinley1993, Arratia2006,Sousa2015, Sousa2018} and numerical \cite{Deiber1981, Poole2007, Grilli2013} efforts have characterized the onset and impact of elastic instabilities in well-defined geometries including cross slot ~\cite{Arratia2006, Poole2007}, Couette~\cite{Larson1990, Groisman1998}, Poiseuille~\cite{Groisman2001, Meulenbroek2003}, and ordered pillar array flows~\cite{McKinley1993, Arora2002, Qin2017}. 
In contrast to the high degree of symmetry in such systems, how geometrical disorder affects the onset of elastic instability remains an open question.
\par

\begin{figure*}[t]
\includegraphics[width=1\textwidth]{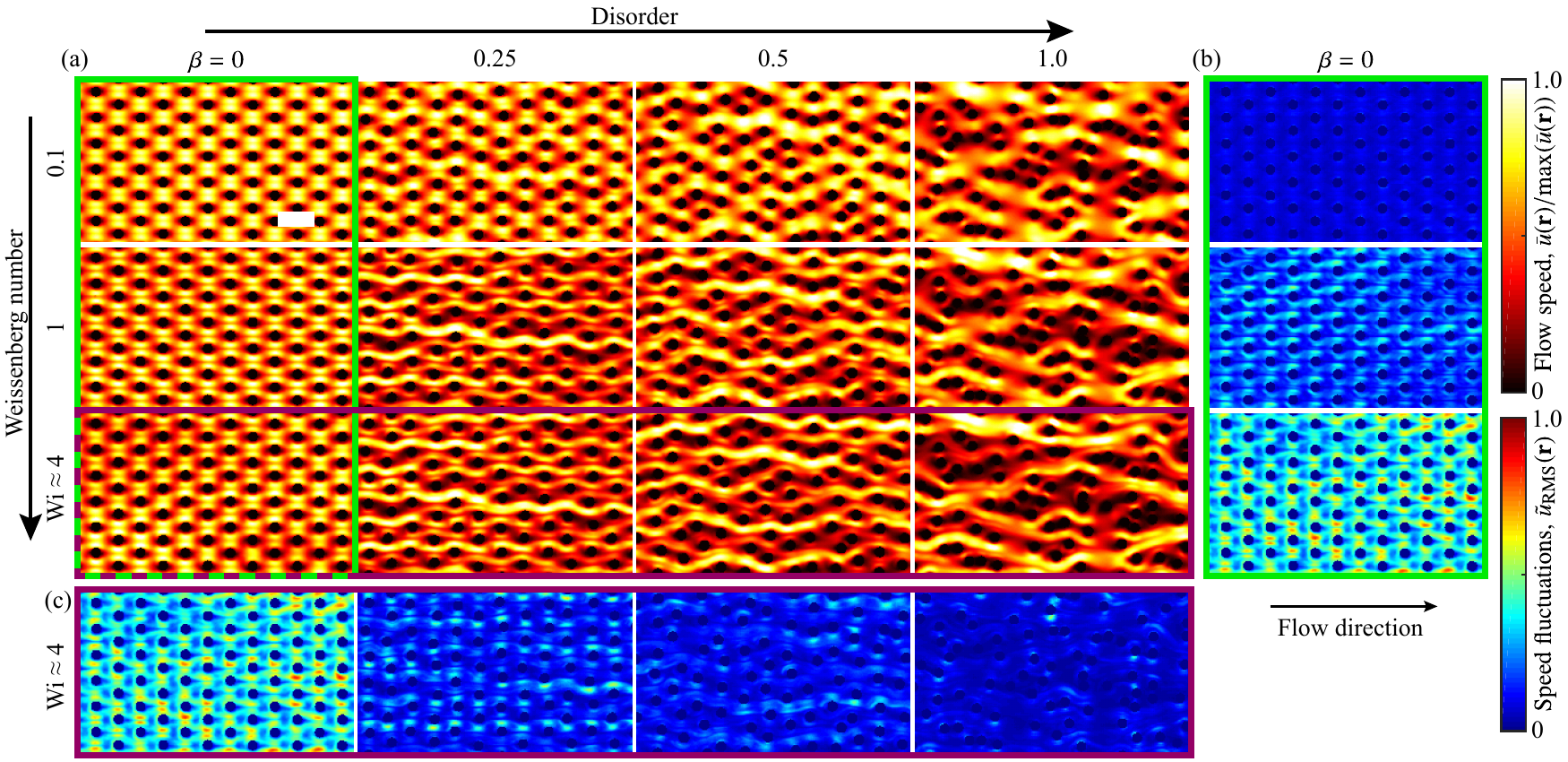}
\vspace{-2em}
\caption{Disorder reduces chaotic fluctuations in viscoelastic flows. 
(a) Normalized, time-averaged speed field, $\bar{u}(\mathbf{r}) / \max(\bar{u}(\mathbf{r}))$, in a microfluidic pillar array for a range of Weissenberg numbers, $\mathrm{Wi}$, and geometric disorders, $\beta$ (40\% of full field of view shown; see also Supplemental Movies~1-5~\cite{SIRef}). Scale bar, 150~$\mu$m. 
(b) Local, normalized speed field fluctuations, $\tilde{u}_{\textrm{RMS}}(\mathbf{r})$, as a function of increasing $\mathrm{Wi}$, corresponding to speed fields for $\beta=0$ [outlined in green in (a)]. (c) Local, normalized speed fluctuations as a function of increasing disorder, corresponding to speed fields for $\mathrm{Wi}\approx 4$ [outlined in magenta in (a)]. 
\label{Fig1}}
\vspace{-2em}
\end{figure*}

Geometrical disorder is a fundamental determinant of transport properties for diverse physical systems, ranging from Anderson localization~\cite{Anderson1958} to colloidal glasses~\cite{Weeks2002} to network dynamics~\cite{Watts1998}.
Similar to viscoelastic flows, coupled dynamical systems are known to display chaotic dynamics under sufficient driving force~\cite{Shinbrot1992}.
However, simulations have suggested that disorder can promote synchronization among arrays of oscillators~\cite{Braiman1995} and in earthquakes~\cite{Mousseau1996}, a phenomenon which has been realized in relatively few physical systems~\cite{Sagues2007,Shew1999}.
In this Letter, we experimentally demonstrate the role of geometric disorder in inhibiting the transition to chaos for viscoelastic flows through microfluidic pillar arrays (Fig.~\ref{Fig1}). 
In an ordered pillar array, the flow field fluctuations undergo a bifurcation at the predicted $\mathrm{Wi}\approx 0.5$~\cite{DeGennes1974}.
Strikingly, the introduction of small, finite disorder to the pillar array (12.5\% of the lattice constant) delays the onset of velocity fluctuations to higher $\mathrm{Wi}\approx 1.2$, and larger disorders ($\ge25\%$) completely quench the transition.
This work thus provides a rare, physical realization of a novel class of systems, where disorder suppresses chaos.\par

Microfluidic channels (25~mm long, 4~mm wide, 50~$\mu m$ high) containing arrays of cylindrical pillars (diameter, $d=50$~$\mu$m) were fabricated using soft lithography to provide precise control over the geometry. 
Starting from a hexagonal array (lattice constant, $a=120$~$\mu$m), disorder is introduced into the pillar arrangements through computer-generated photolithography masks.
Pillar locations are randomly displaced from the ordered lattice within a hexagon of circumradius, $\beta a$ (see Supplemental Material~\cite{SIRef}), where five individual microchannels were fabricated with disorders of $\beta =$~[0, 0.125, 0.25, 0.5, 1.0].

The viscoelastic fluid is comprised of high molecular weight polyacrylamide (PAA; $18\times10^6$~g/mol) at 150~ppm in a viscous Newtonian solvent (97\% aqueous glycerol), which ensures a dilute polymer regime (overlap concentration, $c^\ast=350$~ppm)~\cite{Arratia2006}.
The shear rheology (TA AR-2000) of the PAA solution exhibits a slightly shear thinning viscosity, $\eta$, in the range 2~Pa-s~$\ge \eta \ge$~0.5~Pa-s for shear rates $0.01 \le \dot{\gamma} \le 10$~s$^{-1}$.
The elastic relaxation time of the fluid, $\tau=1.14\textrm{ s}\pm0.1$~s, was measured using a capillary breakup extensional rheometer (CaBER)~\cite{SIRef}. 
The viscoelastic fluid is driven through the microfluidic pillar arrays at constant pressure (Elveflow OB1), and video microscopy (Nikon Ti-e; 10$\times$, 0.2 NA objective) captures the motion (100 fps; Andor Zyla) of fluorescent tracer particles (diameter, $0.5$~$\mu$m) seeded in the flow. 
Time-resolved velocity fields, $\textbf{u}(\textbf{r},t)$, are measured using particle image velocimetry (PIV)~\cite{Thielicke2014}, and Lagrangian flow statistics are obtained by simultaneously tracking slightly larger particles (diameter, $1$~$\mu$m) seeded at low density.
The maximum Reynolds number across experiments is $\mathrm{Re} = \rho U d /\eta \lesssim 10^{-4}$ (density, $\rho$; mean flow speed, $U$), ensuring that inertial effects are negligible.\par

The introduction of geometric disorder into the pillar arrays shifts the flow topology from highly periodic to heterogeneous and decreases the temporal fluctuations of the velocity field.
Flow speed fields, $u(\textbf{r},t) =|\textbf{u}(\textbf{r},t)|$, are time averaged, $\bar{u}(\textbf{r}) = \langle u(\textbf{r},t)\rangle_{t}$, to quantify flow topology [Fig.~\ref{Fig1}(a)] as a function of both disorder, $\beta$, and flow strength, where the latter is characterized by the Weissenberg number, $\textrm{Wi}=\tau U/d$. 
At low $\mathrm{Wi}$, disorder induces heterogeneities in the time-averaged speed field [Fig.~\ref{Fig1}(a), $\mathrm{Wi} \approx 0.1$], similar to Newtonian flow~\cite{Alim2017}.
As $\mathrm{Wi}$ is increased, the flow speed in fast flowing regions becomes amplified leading to `channelization' of the flow field [Fig.~\ref{Fig1}(a), $\beta=1.0$]. 
In ordered geometries, the measured local temporal fluctuations~\cite{Arratia2006}, $\tilde{u}_{\textrm{RMS}}(\textbf{r}) = \sqrt{\langle (\tilde{u}(\textbf{r},t)-\langle\tilde{u}(\textbf{r},t)\rangle_t)^2\rangle_t}$, of the normalized speed field, $\tilde{u}(\textbf{r},t)=u(\textbf{r},t)/U$, increase with $\textrm{Wi}$, as expected for chaotic regimes with $\textrm{Wi} \gtrsim 1$ [Fig.~\ref{Fig1}(b)].
In surprising contrast, the amplitude of the temporal flow speed fluctuations drastically decreases by an order of magnitude as the disorder of the system is increased at high $\mathrm{Wi}$ [Fig.~\ref{Fig1}(c), $\textrm{Wi} \approx 4$; see also Supplemental Movies~1-5~\cite{SIRef}].
The viscoelastic flows through pillar arrays studied here demonstrate a surprising interplay between geometry, flow topology, and elastic stability.

\begin{figure*}[t]
\includegraphics[width=1\textwidth]{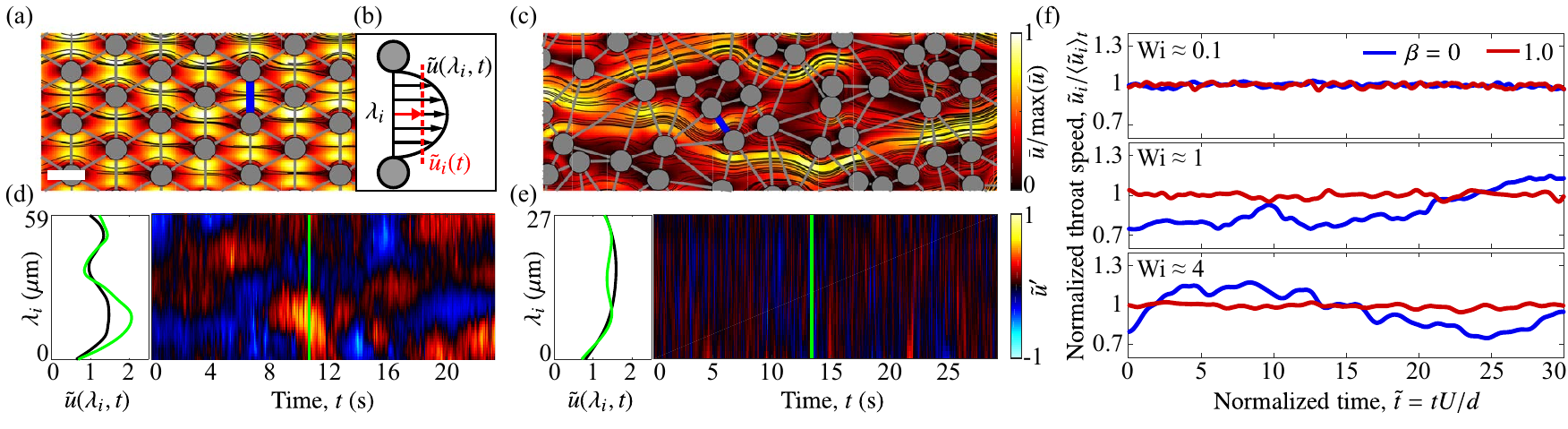}
\vspace{-2em}
\caption{Pore-scale velocity fluctuations are indicative of elastic instability. 
(a) Normalized, time-averaged speed field for the ordered lattice ($\beta = 0$, $\mathrm{Wi} \approx 4$).  Streamlines (black) are computed from measured flow fields, and pillar locations (gray circles) are used to discretize throat cross sections (gray lines). Scale bar, 70~$\mu$m. 
(b) Schematic of the throat flow profile, $\tilde{u}(\lambda_i,t)$, and spatially-averaged speed, $\tilde{u}_i(t)$, with local coordinate, $\lambda_i$.
(c) Normalized, time-averaged speed field for a disordered lattice ($\beta = 1$, $\mathrm{Wi} \approx 4$). 
(d)-(e) Measured instantaneous ($\tilde{u}(\lambda_i,t)$; green) and time-averaged ($\langle \tilde{u}(\lambda_i,t)\rangle_t$; black) throat speed profiles (left) and flow speed fluctuation, $\tilde{u}_i^\prime=\tilde{u}_i(t)-\langle\tilde{u}_i(t)\rangle_t$, kymographs (right) for individual throats from (d) ordered and (e) disordered channels [blue throats in (a) and (c), respectively; $\mathrm{Wi}\approx4$]. 
Instantaneous speed profiles (left, green) correspond to indicated kymograph time (right, green). 
(f) Normalized, instantaneous throat flow speed for disordered ($\beta = 0$) and ordered ($\beta = 1$) throats at three different $\mathrm{Wi}$ values [blue throats in (a) and (c), respectively].
\label{Fig2}}
\vspace{-1.5em}
\end{figure*}

The spatial extent of the velocity fluctuations is typically on the order of the pillar spacing (Fig.~\ref{Fig2}), which suggests that pore-scale averaging of the flow statistics provides an adequate level of course graining to study the dynamical transition of these viscoelastic flows~\cite{Pan2013}.
Taking a hydraulic network point of view, we inspect the fluctuations within throats between adjacent pillars [Figs.~\ref{Fig2}(a)-(c)] using a Delaunay triangulation of their positions.
The normalized speed fields are interpolated [Fig.~\ref{Fig2}(b)] between pillars to obtain a time-dependent speed profile [Fig.~\ref{Fig2}(d)-(e)], $\tilde{u}(\lambda_i,t)$, where $\lambda_{i}$ runs across throat $i$.
Kymographs reveal deficits and excesses of flow speed in a given throat cross section [Fig.~\ref{Fig2}(d)-(e)] and are used to discriminate sub-pore-scale temporal fluctuations [Fig.~\ref{Fig2}(d)].
These fluctuations are comparable to the pore scale, and thus, we compute the instantaneous, normalized mean flow speed through each throat, $\tilde{u}_i(t) = \langle \tilde{u}(\lambda_i,t)\rangle_{\lambda_i}$.
The normalized throat flow speed initially exhibits small fluctuations in the ordered system, which markedly grow with increasing $\mathrm{Wi}$ [Fig.~\ref{Fig2}(f), $\beta=0$].
In distinct contrast, throat speeds for the disordered system remain nearly constant across $\mathrm{Wi}$ [Fig.~\ref{Fig2}(f), $\beta=1$] with comparable fluctuations to the $\mathrm{Wi}\approx 0.1$ ordered system [Fig.~\ref{Fig2}(f), $\beta=0$].
This pore-scale analysis captures the essential features of the system (Fig.~\ref{Fig1}) and provides a convenient framework to determine how disorder affects the dynamical transition to chaos.
\par

\begin{figure}
\includegraphics{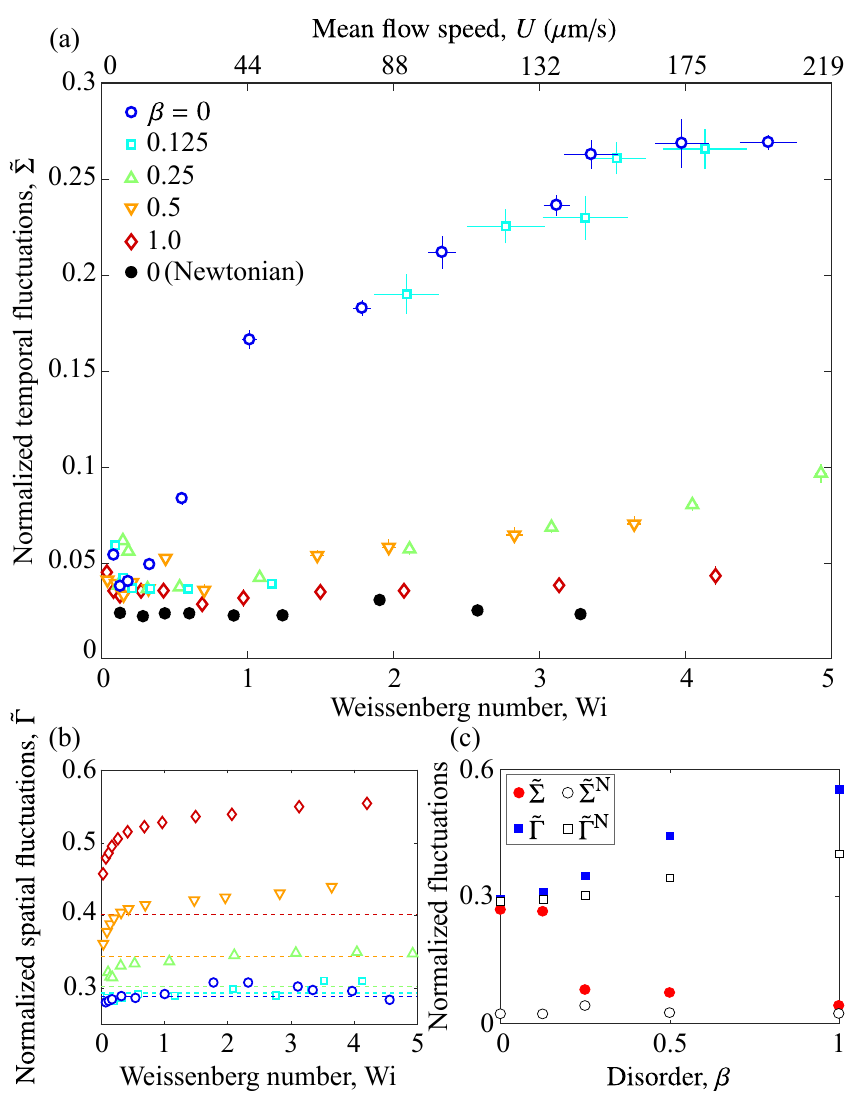}
\vspace{-2em}
\caption{Disorder delays the onset of viscoelastic instability. 
(a) Normalized temporal fluctuations of throat flow speed for viscoelastic fluid flow (open symbols) as a function of Weissenberg number, $\mathrm{Wi}$, for a range of geometric disorders, $\beta$. Temporal fluctuations for a Newtonian fluid flow ($\beta=0$; filled black circles) are shown for reference (see upper horizontal axis). 
(b) Normalized spatial fluctuations of the time-averaged throat flow speed for a viscoelastic fluid as a function of $\mathrm{Wi}$ in various geometric disorders. Dotted horizontal lines represent measured Newtonian values. 
(c) Temporal and spatial fluctuations of both viscoelastic and Newtonian fluids for various $\beta$ at fixed mean flow speed corresponding to $\textrm{Wi}\approx4$.
\label{Fig3}}
\vspace{-1.5em}
\end{figure}

Examination of the pore-scale flow speed fluctuations demonstrates that the onset of time-dependent flow undergoes a global, forward bifurcation~\cite{Pan2013} in the ordered geometries but not in disordered geometries.
The ensemble-averaged, temporal fluctuation of the normalized throat speeds, $\tilde{\Sigma} = \langle \sqrt{\langle(\tilde{u}_i(t)-\bar{u}_i)^2\rangle_t} \rangle_{i} $, is used as the order parameter, where $\bar{u}_i=\langle\tilde{u}_i(t)\rangle_t$ [Fig.~\ref{Fig3}(a)].
For the ordered lattice ($\beta=0$), a bifurcation in the throat speed fluctuations occurs at $\textrm{Wi}_\textrm{cr}\approx 0.5$, which is accompanied by the onset of slow flow speed fluctuations (see Supplemental Movie~2~\cite{SIRef}) and corresponds to the predicted polymer coil-stretch transition in purely extensional flows~\cite{DeGennes1974}. 
A minor perturbation to the ordered geometry ($\beta=0.125$) significantly delays the transition to $\textrm{Wi}_{\mathrm{cr}}\approx 1.2$, where time-dependent flow occurs.
At yet higher disorders, the flow appears stable up to the limits of our experiment ($\textrm{Wi} \approx 5$), where for $\beta\geq 0.25$ no bifurcation is discernible and speed fluctuations are significantly damped. 
For example, $\tilde{\Sigma}$ for viscoelastic flows with $\beta=1.0$ is one order of magnitude smaller than for $\beta = 0$ and comparable to $\tilde{\Sigma}$ for Newtonian flow at similar mean speeds [Figs.~\ref{Fig3}(a)~and~\ref{Fig3}(c)].
To our knowledge, this result represents the first observation of the stabilizing effect of disorder on viscoelastic flow. 
Furthermore, the finite disorder necessary to quench the onset of time-dependent flow and lack of scaling with $\textrm{Wi}$ across disorder, suggests a deeper coupling between flow topology and polymer stretching.

Disorder suppresses the abrupt onset of temporal fluctuations, but introduces spatial heterogeneity into the flow topology, where viscoelasticity further enhances flow channelization~\cite{SIRef} in a continuous fashion.
The normalized, spatial throat speed fluctuations - defined by the root-mean-square, $\tilde{\Gamma} =\sqrt{\langle(\bar{u}_i-\langle \bar{u}_i\rangle_{i})^2\rangle_{i}}$ - reveal a continuous increase with $\textrm{Wi}$ from their respective, measured Newtonian values, $\tilde{\Gamma}^N$, for highly disordered geometries ($\beta = [0.25, 0.5, 1.0]$) [Fig.~\ref{Fig3}(b)]. 
In contrast, ordered geometries ($\beta = [0, 0.125]$), where bifurcations in $\tilde{\Sigma}$ are evident, show relatively little change in $\tilde{\Gamma}$ with  $\textrm{Wi}$ [Fig.~\ref{Fig3}(b)].
The result is an apparent trade-off between spatial and temporal fluctuations with increasing disorder [Fig.~\ref{Fig3}(c)].
The non-linear flow response of disordered systems to increased $\mathrm{Wi}$ goes beyond a simple increase in the spatial heterogeneity of flow speed, and the observed channelization [Figs.~\ref{Fig1}(a)~and~\ref{Fig2}(c)]~\cite{Datta2013} likely alters the nature of the deformations experienced by fluid particles.

\begin{figure}
\includegraphics{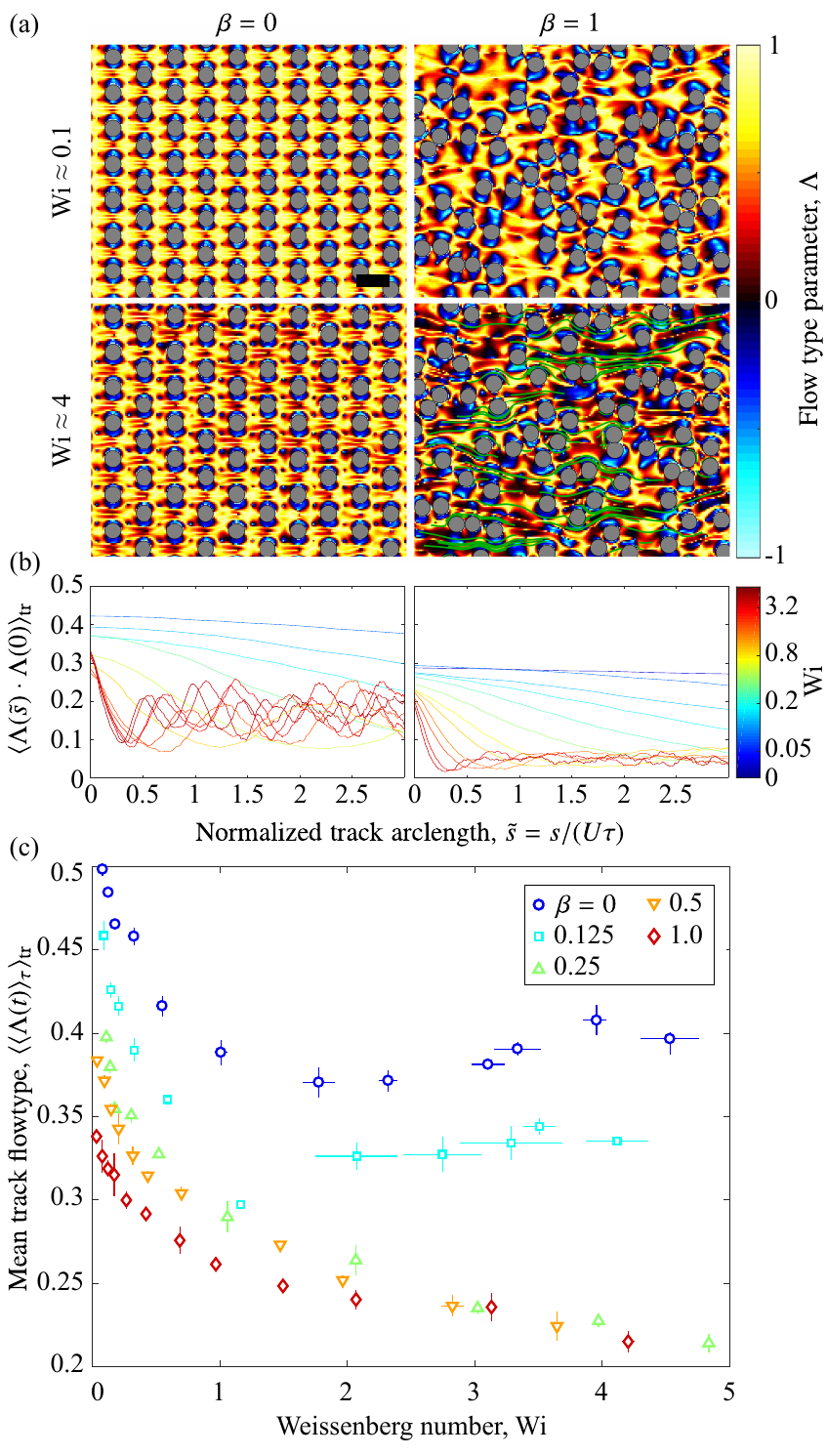}
\vspace{-2.5em}
\caption{Disorder regulates extensional flow type history and thus the transition to chaotic flow.
(a) Flow type parameter maps measured from time-averaged velocity fields for ordered and disordered channels at low and high Wi. Particle trajectories (green) are shown for one relaxation time, $\tau$ (lower right). Scale bar, 100~$\mu$m.
(b) Auto-correlation of the local flow type along tracer particle trajectories for $\beta=0$ (left) and $\beta=1$ (right) across $\textrm{Wi}$. 
(c) Ensemble-averaged flow type along tracer particle trajectories over one polymer relaxation time, $\tau$, as a function of $\textrm{Wi}$ for all geometries.
\label{Fig4}}
\vspace{-2em}
\end{figure}

The local mode of fluid deformation, or flow type, dramatically affects the hydrodynamic response of viscoelastic fluids. 
For example, in extensional strain, viscoelastic flows reach a critical stretching rate, $\dot{\epsilon}=(2\tau)^{-1}$, due to finite polymer extensibility, accompanied by increased extensional viscosity~\cite{DeGennes1974}. 
The flow type parameter~\cite{Astarita1979}, $\Lambda=(||\mathbf{D}||-||\mathbf{\Omega}||)/(||\mathbf{D}||+||\mathbf{\Omega}||)$, quantifies the local flow kinematics ranging from pure rotation ($\Lambda=-1$) to shear ($\Lambda=0$) to pure extension ($\Lambda=+1$), where $\mathbf{D}$ and $\mathbf{\Omega}$ are the strain-rate and vorticity tensors, respectively, and $||\mathbf{D}||=\sqrt{2\mathbf{D}:\mathbf{D}}$. 
At low $\mathrm{Wi}$, the flow through both ordered ($\beta=0$) and disordered ($\beta = 1$) arrays is dominated by extension [Fig.~\ref{Fig4}(a), top row].
As $\textrm{Wi}$ increases, the flow type in the ordered geometry remains primarily extensional [Fig.~\ref{Fig4}(a), bottom left]. 
However, a clear shift toward shear-dominated flow-type is evident in the disordered geometry [Fig.~\ref{Fig4}(a), bottom right], suggesting that the flow type experienced by fluid particles is integral to the stabilizing mechanism of these viscoelastic flows.

The degree of polymer stretching is dependent upon the Lagrangian flow type history of fluid particles, which ultimately dictates the global dynamics of viscoelastic flows~\cite{Wagner2015}.
The auto-correlation of the flow type [Fig. \ref{Fig4}(b)] along measured particle trajectories [Fig. \ref{Fig4}(a), bottom right, green tracks] quantifies the constancy of fluid deformation.
In ordered geometries [Fig. \ref{Fig4}(b), left, $\beta=0$], fluid particles are subjected to strongly extensional flow with regular frequency, whereas the weakly correlated flow type experienced in random media likely facilitates polymer relaxation [Fig.~\ref{Fig4}(b), right, $\beta=1$] despite the spatial correlations introduced by flow channelization (Fig.~\ref{Fig1}).
To compare the relative flow type experienced by fluid particles across geometries, we compute the ensemble-averaged, mean flow type (see Fig.~S8~\cite{SIRef}) along measured particle trajectories [Fig.~\ref{Fig4}(a), bottom right] over one relaxation time, $\tau$ [Fig.~\ref{Fig4}(c)].
This time-averaged flow type initially decreases with $\textrm{Wi}$ for all disorders, tending toward shear. 
However, the most ordered geometries ($\beta=[0,0.125]$) plateau at $\Lambda > 0.3$ for $\textrm{Wi} \gtrsim 1$, corresponding to the unstable regime. 
These results show that disordered media enable viscoelastic flow to minimize extensional strain and avoid elastic instability, whereas ordered microstructure maintains sufficiently strong, coherent extension to trigger instability.

Lagrangian unsteady flows, having non-constant stretch history, present theoretical challenges for understanding elastic flow instabilities~\cite{Wagner2015}. 
However, we can gain some insight into the geometry-dependent conditions, $M_\textrm{cr}$, for elastic stability through the well-developed Pakdel-McKinley criterion~\cite{Pakdel1996,McKinley1996}:
\begin{equation}
\left[\frac{\tau U}{\mathcal{R}}\frac{\sigma_{11}}{\eta_0 \dot{\gamma}} \right]^{1/2} \geq M_{\textrm{cr}},
\end{equation}
where $\mathcal{R}$ is the radius of curvature of a streamline, $\sigma_{11}$ is the stream-wise tensile stress, and $\eta_0$ is the zero shear rate viscosity. 
The first term, $\tau U/\mathcal{R}$, represents the contribution of geometry to polymer stretching through streamline curvature, but this effect has only a minor variation across disorders (see Fig.~S5 in the Supplemental Material~\cite{SIRef}).
The second term in the criterion, $\sigma_{11}/(\eta_0\dot{\gamma})$, is the ratio of extensional to shear stresses. 
While we do not have access to the local $\sigma_{11}$, the observed evolution to shear-dominated flow with increasing $\mathrm{Wi}$ and disorder (Fig.~\ref{Fig4}) is consistent with reducing the Pakdel-McKinley number, tending toward stable flow.

In this Letter, we demonstrate that the introduction of finite disorder into hydraulic networks suppresses the onset and amplitude of chaotic velocity fluctuations associated with viscoelastic flow, emphasizing the need for a Lagrangian understanding of elastic flow stability beyond the Weissenberg number~\cite{Pakdel1996,McKinley1996}.
The stabilizing effect of disorder is attributed to a shift in the flow type history of the viscoelastic fluid, which evolves from extensional and unstable in ordered systems to shear-dominated and stable in disordered systems.
We expect that the transport properties of these transitional flows will likewise be sensitive to disorder due to temporal fluctuations, which are important for extraction, remediation, and other industrial processes in porous media flows~\cite{Scholz2014, Clarke2015}.
More broadly, coupled systems often exhibit chaotic dynamics under sufficient forcing, where only a few examples have thus far demonstrated the counterintuitive restoration of synchrony or stability by disorder~\cite{Sagues2007}:
Arrays of forced, coupled pendula desynchronize, but introducing disorder among their natural frequencies suppresses chaotic oscillations and recovers periodicity~\cite{Braiman1995, Shew1999}.
In earthquakes and neural networks, finite disorder causes a transition from a self-organized-critical (SOC) distribution of avalanches to system-wide, periodic events~\cite{Mousseau1996,Fontenele2019}.
Analogously, in the system presented here, stability is maintained for a viscoelastic flow through a disordered microstructure, relative to a perfectly crystalline medium.
Hence, this work provides a new, physical example of the suppression of chaos via disorder and adds to the growing canon of this important phenomenon, which is thus far dominated by simulation and theory.

\begin{acknowledgments}
We thank G.H. McKinley and J. Du for assistance with CaBER and shear rheology, and P.E. Arratia and D.A. Gagnon for helpful discussions. 
This work was supported by National Science Foundation awards CBET-1701392, CAREER-1554095, and CBET-1511340 (to J.S.G.).
\end{acknowledgments}


%




\end{document}